\documentclass[prl,twocolumn,showpacs,preprintnumbers]{revtex4}

\usepackage{epsfig}
\usepackage{graphicx}
\usepackage{bm}

\DeclareGraphicsExtensions{.eps}
\begin{document}
\title{Direct Production of Tripartite Pump-Signal-Idler Entanglement in the Above-Threshold Optical 
Parametric Oscillator}

\author{A. S. Villar$^1$}
\author{M. Martinelli$^1$}
\author{C. Fabre$^2$}
\author{P. Nussenzveig$^1$}
\email{nussen@if.usp.br}
\affiliation{$^1$Instituto de F\'\i sica, Universidade de S\~ao Paulo, Caixa Postal 66318, 
05315-970 S\~ao Paulo, SP, Brazil \\ 
$^2$Laboratoire Kastler Brossel, Case 74, Universit\'e Pierre et Marie
Curie - Paris 6, 4 Place Jussieu, 75252 Paris Cedex 05, France}

\date{\today}
\begin{abstract}
We calculate the quantum correlations existing among the three
output fields (pump, signal, and idler) of a triply resonant
non-degenerate Optical Parametric Oscillator operating above
threshold. By applying the standard criteria [P. van Loock and A. Furusawa, 
Phys. Rev. A \textbf{67}, 052315 (2003)], we show that strong tripartite 
continuous-variable entanglement is present in this well-known and simple system. 
Furthermore, since the entanglement is generated directly from a nonlinear 
process, the three entangled fields can have very different frequencies, opening 
the way for multicolored quantum information networks. 
\end{abstract}



\maketitle

Entanglement, which is probably the strangest of all quantum phenomena, is 
considered the most important resource for future quantum information technology. 
Recent experiments on quantum computing, storage and communication of 
information utilize different ``quantum hardware'', such as atom 
clouds~\cite{felintonature05}, quantum dots~\cite{imamogluscience06} and 
trapped ions~\cite{multipartiteionsnature}, all with different resonance 
frequencies. These systems will probably be used in nodes of quantum networks, 
implying the necessity of devising ways to address them without loss of 
quantum information. For networks with several nodes, multipartite entangled 
light beams will be important to carry out such tasks. 

Most of the current realizations of entangled light beams are implemented 
by combining squeezed beams on beam splitters~\cite{Teleport,LeuchsEPR,
LoockBraunstein,Aoki3part,AokiNature,Peng3part}. The beam splitter 
transformation is linear and does not lead to entangled beams of different 
frequencies. In order to produce multicolored entangled beams it is 
important to generate them directly from a nonlinear process. In the 
case of bipartite two-color entanglement, this has been done very recently, 
in the above-threshold optical parametric oscillator 
(OPO)~\cite{entOPO,optlettpeng,pfisterentang}. 

The OPO is the best known and most widely used source of entangled 
continuous variables for quantum information purposes~\cite{loockbraunrmp}. Nevertheless, 
focus thus far has been on the down-converted beams it produces, 
usually overlooking quantum properties of the pump beam. Recent proposals for direct 
generation of tripartite entanglement use so-called cascaded nonlinearities, 
combining down-conversion and sum/subtraction frequency 
generation~\cite{tricascade}, which are not present in standard OPOs. In 
this Letter, we theoretically demonstrate that the standard triply resonant 
above-threshold OPO naturally produces pump-signal-idler tripartite entanglement. 
We show that the down-converted and the pump fields' noises violate 
inequalities which are sufficient for witnessing 
entanglement~\cite{loockbraunrmp}. We believe this to be the simplest 
and most practical proposal of a multicolored multipartite entanglement 
source to date. 

For tripartite systems with subsystems $(k, m, n)$, if the state
is partially separable, the density operator can be written in the 
form of a statistical mixture of reduced density operators 
$\hat\rho_{i,km}$ and $\hat\rho_{i,n}$: 
\begin {equation}
\hat\rho=\sum_i \eta_i \hat\rho_{i,km}\otimes\hat\rho_{i,n}\;,
\label{separa}
\end{equation}
with weights $\eta_i\geq 0$ satisfying $\displaystyle \sum_i\eta_i 
=1$. A necessary condition for separability of two 
sub-systems was demonstrated by Duan \textit{et al.}~\cite{DGCZ}, 
in the form of an inequality: if it is violated, there is 
bipartite entanglement. This criterion is easily checked 
experimentally by measuring second order moments of 
combinations of operators acting on each of the subsystems. 

The inequality presented in Ref.~\cite{DGCZ} for the variances of 
two combinations of positions and momenta $(\hat x_j,\hat p_j)$ of 
subsystems $j=\{1,2\}$ can be readily extended to a 
combination of three subsystems~\cite{LoockFuru}. If we define two 
commuting operators $\hat u=h_1\hat x_1 +h_2\hat x_2 +h_3\hat x_3$ 
and $\hat v=g_1\hat p_1 +g_2\hat p_2 +g_3\hat p_3$, where the $h_i$ and 
$g_i$ are arbitrary real parameters, for a (partially) separable 
state written in the form of eq.~(\ref{separa}), inequalities 
of the form:
\begin {equation}
\langle \Delta^2\hat u\rangle+\langle \Delta^2\hat v\rangle\geq 
2(|h_n g_n|+|h_k g_k+h_m g_m|) \;,
\label{LF}
\end{equation}
with different permutations of the subsystems $(k,m,n)$, must hold. 
Therefore, violations of the inequalities corresponding to the 
three possible permutations suffice to demonstrate tripartite entanglement. 

For electromagnetic fields, position and momentum operators can be replaced by the
field amplitude and phase quadrature operators, defined as functions of the creation and
annihilation operators as
$\hat p_j(t)=[e^{i\varphi_j} \hat a_j^\dag(t)+e^{-i\varphi_j} \hat a_j(t)]$ and
$\hat q_j(t)=i[e^{i\varphi_j} \hat a_j^\dag(t)-e^{-i\varphi_j} \hat a_j(t)]$,
where the phase $\varphi_j$ of each mode is chosen from its mean value in order to have
$\langle \hat q_j\rangle = 0$. In this case, $\hat p$ represents the amplitude fluctuations of
the field, and $\hat q$ is related to the phase fluctuations. From the commutation relation
$[\hat a_j, \hat a_{j'}^\dag]=\delta_{jj'}$, it follows that
$[ \hat p_j,  \hat q_{j'}]=2i\delta_{jj'}$.
In the present situation, we look for violations of the following inequalities:
\begin{eqnarray}
S_1=\langle \Delta^2(\hat p_1-\hat p_2)\rangle + \langle \Delta^2(\hat q_1+\hat q_2-\alpha_0 \hat q_0)\rangle \geq 4 ,\label{crit1}\\
S_2=\langle \Delta^2(\hat p_0+\hat p_1)\rangle + \langle \Delta^2(\hat q_1+\alpha_2 \hat q_2-\hat q_0)\rangle \geq 4 ,\label{crit2}\\
S_3=\langle \Delta^2(\hat p_0+\hat p_2)\rangle + \langle \Delta^2(\alpha_1 \hat q_1+\hat q_2-\hat q_0)\rangle \geq 4 ,\label{crit3}
\end{eqnarray}
with an optimized choice of the free parameters $\alpha_i$, in 
order to show that all three modes are entangled, i.e. that the 
state of the full system is not even partially separable.

The tripartite entangled fields are directly produced by a triply resonant 
nondegenerate optical parametric oscillator (OPO), composed of a $\chi^{(2)}$ 
nonlinear crystal placed inside an optical cavity (a full 
description of field mean values and tuning characteristics can be 
found in Ref.~\cite{Debuisschert}). The OPO is a well-known source 
of non\-classical states of the electromagnetic field, both above 
and below the oscillation threshold. In this system, a pump photon 
of frequency $\omega_0$ is down-converted into a pair of signal 
and idler twin photons of frequencies $\omega_1$ and $\omega_2$. 
These fields exit the cavity and can be easily separated by 
color (pump) and polarization (signal and idler) in the case of 
type-II phase matching. Below threshold, signal and idler modes 
are in an entangled state with zero mean values for the electric 
field~\cite{PengEPR}. Above threshold, the parametric 
coupling leads to both intensity coupling between the three modes 
(this is the well-known pump depletion effect: a pump photon is 
destroyed each time a couple of twin and idler photons is created) 
and to phase coherence between them: the sum of the signal and 
idler field phases is locked to the pump phase as a consequence of 
energy conservation ($\omega_1+\omega_2=\omega_0$). This leads to 
both intensity and phase correlations between the three modes that 
extend to the quantum regime, and eventually culminate in 
tripartite entanglement as we show below. So far, physicists' interest has 
been concentrated on the signal and idler quantum 
correlations~\cite{Camy} or on the pump squeezing~\cite{Kasai}. The full 
three-mode system has indeed genuine quantum properties~\cite{drummtriplecorrel}, 
which are partly lost when one traces out the pump mode, although the signal 
and idler modes remain of course entangled~\cite{entOPO}.

Quantum fluctuations of the system are calculated as usual~\cite{Fabre}: 
we start from the evolution equations of the operators of the three modes 
$(\hat a_0,\hat a_1,\hat a_2)$ inside the OPO cavity. We write the field operators 
as the sum of their mean values and a fluctuation term and, 
assuming that the fluctuations are small compared to the mean 
fields, which is true everywhere except very close to threshold, 
we linearize these equations around the classical mean 
values~\cite{Debuisschert}. One obtains in this way six linear
Langevin equations that enable us to calculate the evolution of
the real and imaginary parts of the intracavity fluctuations of
the three fields. If we assume that the cavity transmission factor
and the extra-losses are the same for the signal and idler modes,
the evolution equations can be decoupled into two independent
sets~\cite{Fabre}: two equations for the signal and idler
difference, and four equations coupling the sum of the signal and
idler fluctuations to the pump fluctuations. Using the
input-output relation on the coupling mirror, one obtains the
output field fluctuations in Fourier domain, 
$\delta \vec{p}(\Omega) = [\delta \hat p_0(\Omega), 
\delta \hat q_0(\Omega), \delta \hat p_1(\Omega), 
\delta \hat q_1(\Omega), \delta \hat p_2(\Omega), 
\delta \hat q_2(\Omega)]^{\mathrm T}$, as a function of the
input field fluctuations. This enables us to determine the full
$6\times6$ three-mode covariance matrix, $C = \langle 
\delta \vec{p}(\Omega) \, \delta \vec{p}(-\Omega)^{\mathrm T} 
\rangle$, of the pump, signal and idler output modes, and the variance of any 
combination of these modes. The full treatment is described in 
Ref.~\cite{VillarOptComm}.

From the calculated covariances, we derive the optimized values of
the parameters $\alpha_i$ which minimize the quantities $S_1$,
$S_2$ and $S_3$ of Eqs. (\ref{crit1}) -- (\ref{crit3}) as
functions of the covariance matrix elements for the output field,
and calculate the corresponding minimum value for these three
quantities. We take typical experimental conditions: cavity
coupling mirror transmittance for pump $T_0=10\%$ and signal and
idler beams $T=2\%$, and exact cavity resonance for the three modes.
We can now study the dependence of $S_1$, $S_2$, and $S_3$ with the
normalized pump power $\sigma$ (power normalized to the
oscillation threshold on resonance) and with the analysis
frequency $\omega$ (normalized to the inverse of the cavity round
trip time $\tau$).

\begin{figure}[ht]
\centering
\epsfig{file=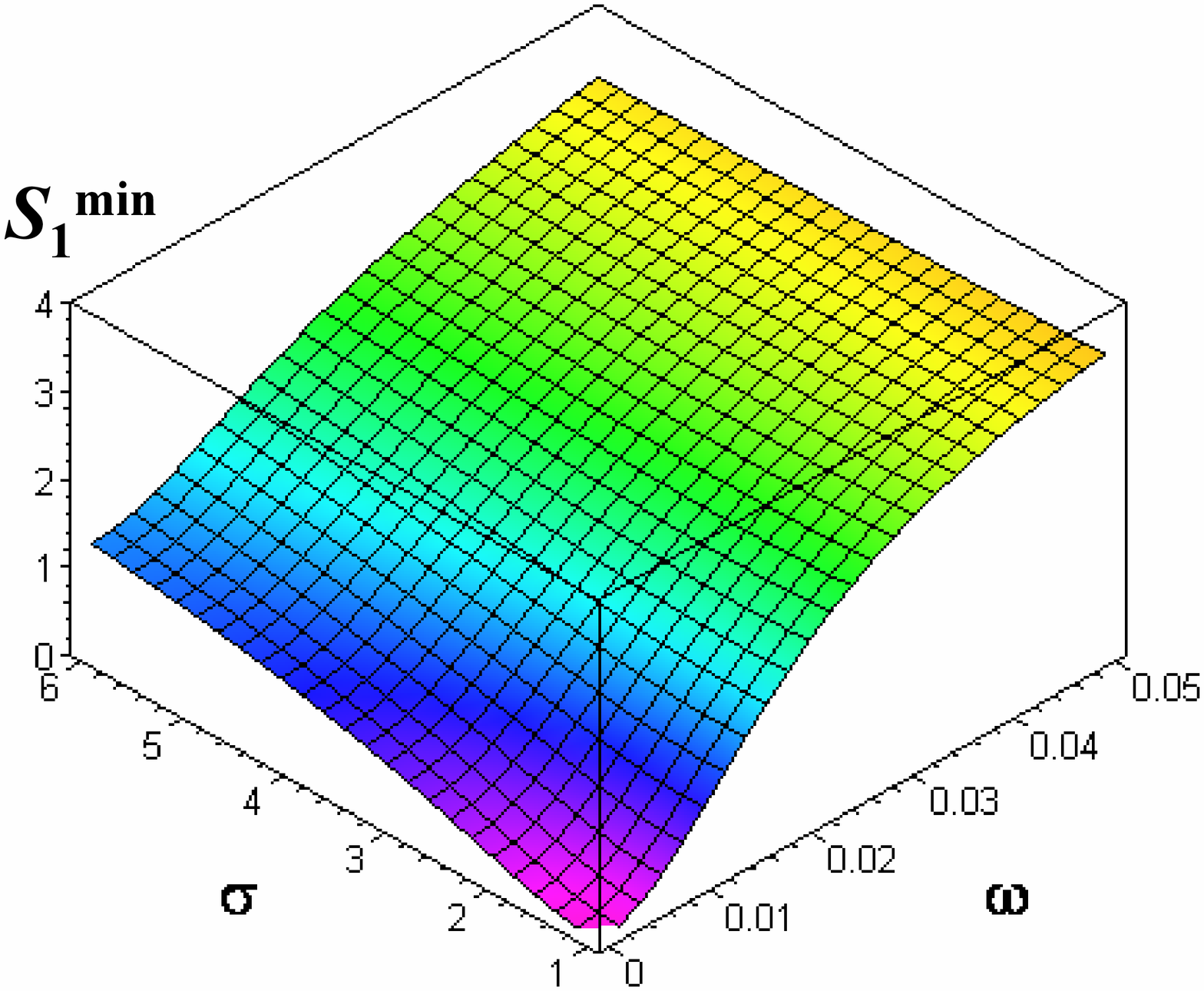,width=70mm}
\caption{\label{crit1o} (color online). Optimized sum of variances, 
$S_1^{\mathrm{min}}$, for eq.~(\ref{crit1}): $\sigma$=pump power relative
to threshold, $\omega$= analysis frequency, in units of 1/$\tau$.}
\end{figure}

\begin{figure}[ht]
\centering
\epsfig{file=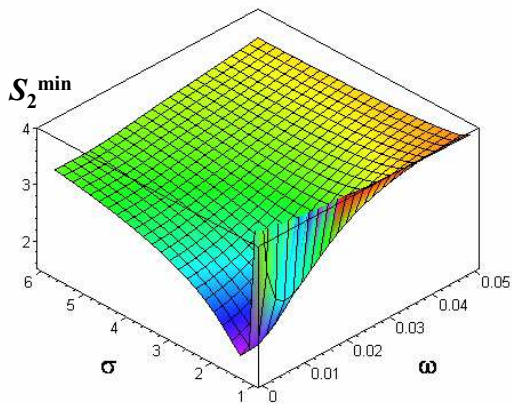,width=70mm}
\caption{\label{crit2o} (color online). Optimized sum of variances, 
$S_2^{\mathrm{min}}=S_3^{\mathrm{min}}$, for eqs.~(\ref{crit2}) and 
(\ref{crit3}): $\sigma$=pump power relative to threshold, $\omega$= 
analysis frequency, in units of 1/$\tau$.}
\end{figure}

In Fig.~\ref{crit1o}, we display the minimized value of $S_1$. As can be
seen, $S_1^{\mathrm{min}}$ is smaller than 4 in all the presented range of
parameters, which establishes the inseparability of the signal and
idler modes. Let us stress that the resulting violation, with the
optimization of the variance, is much stronger than that observed
by tracing out the pump mode and looking only at signal and idler
modes under the Duan {\it et al.} criterion~\cite{VillarOptComm}.
In the present case, the measurement of pump phase increases the
knowledge that one can obtain about the idler beam phase from the
measurement of the signal phase. Nevertheless, the state can still
be partially separable if the other two inequalities
(eqs.~\ref{crit2} and \ref{crit3}) are not violated. The
interchangeability of the roles of signal and idler makes evident
that $S_2=S_3$. The common minimized value of this quantity is
shown in Fig.~\ref{crit2o}. We observe that it is also below 4,
implying inseparability for a broad range of values of analysis
frequency and pump power, even though the violation is not as
strong as for $S_1^{\mathrm{min}}$ (Fig.~\ref{crit1o}). Correlations between the 
twin beams tend to be stronger than those between one of the 
twins and the pump, since the pump is not generated inside the 
cavity. $S_2^{\mathrm{min}}=S_3^{\mathrm{min}}$ is everywhere larger 
than $\simeq 1.7$, a value obtained for $\sigma\simeq1.6$. For this value 
of $\sigma$, all three fields have approximately the same intensities, which 
is in general the best situation for observing correlations.

Another method to characterize the amount of entanglement in a
system is to study the eigenvalues of its covariance matrix: they
provide information about the maximum squeezing that can be obtained
from the different modes by unitary transformations and about the
maximum bipartite entanglement that can be extracted from these
modes~\cite{Adesso}. In our case, the minimum eigenvalue is given
by the variance of $\hat p_1 - \hat p_2$. The next lower eigenvalue is
related to the combination of phases in the form $(\hat q_1+\hat
q_2-\beta \hat q_0)$, where $\beta$ is a real number. Both values 
can be strongly squeezed, at the expense of excess noise for
the variances of their conjugate variables. 

\begin{figure}[ht]
\centering
\epsfig{file=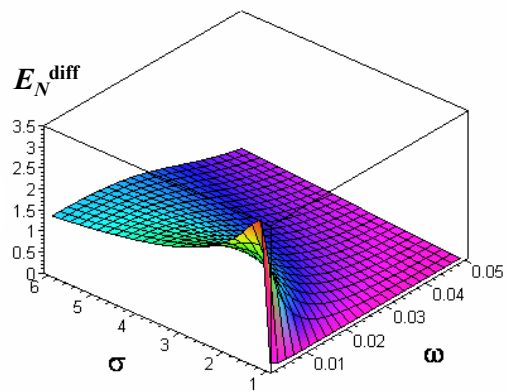,width=70mm}
\caption{\label{En2} (color online). Difference between logarithmic negativities, 
$E_N^{\mathrm{diff}}$, for the full three modes and for only signal and idler modes. 
$\sigma$= pump power relative to threshold, $\omega$= analysis frequency, in units 
of 1/$\tau$.}
\end{figure}

From these two smallest eigenvalues $\lambda_1,\lambda_2$ we calculate 
the logarithmic negativity $E_N=\mathrm{max}[0,-\log_2(\lambda_1 
\lambda_2)/2]$~\cite{Vidal,Wolf}. This is a computable measure of the degree 
of bipartite entanglement of a system, and it is especially useful owing to its 
immediate extension to entangled mixed states. We calculate here the 
difference, $E_N^{\mathrm{diff}}$, between the logarithmic negativities 
for the full system and for just the signal and idler modes, tracing out the 
pump. This difference is positive for the full range of parameters displayed 
in Fig.~\ref{En2}, with maximum values obtained for low analysis frequencies 
($\omega<0.02/\tau$). It is clear that quantum information is present in all 
three modes and one only recovers a fraction of it when restricting measurements 
to signal and idler beams.

The tripartite pump-signal-idler entanglement in the OPO can be 
observed in a broad range of frequencies and pump
power. The correlation is, as expected, stronger for analysis
frequencies below the cavity bandwidth $T/\tau$ for the signal
and idler modes, and for pump powers close to threshold, although
it does not depend so much on this last parameter. Calculations
from the covariance matrix show that there is a small dependence
of $S_1$ and $S_2$ on the cavity detunings, which is important 
because the locking of the OPO is typically done with some small detuning
for pump and down-converted modes. If we consider the presence of
spurious losses inside the cavity, there is a linear increase of
the value of $S_1$ with these losses, much in the way observed for
the intensity correlation of twin beams emitted from the OPO. As
for $S_2$, inseparability no longer occurs for lower 
analysis frequencies, but still holds for a wide range of the
parameters $\sigma$ and $\omega$.

In conclusion, we have demonstrated that the standard nondegenerate 
optical parametric oscillator directly yields 
tripartite entangled light beams when operating above threshold. 
Above-threshold OPOs have produced the highest level of intensity
quantum correlations to date~\cite{Claude10db}. 
Fig.~\ref{crit1o} shows that they can also produce a very low bound 
for the combined phases quantum fluctuations. Thus, the magnitudes of 
expected quantum correlations are among the best achievable at present. The 
experimental realization of this system is much simpler than the proposals 
based on combined nonlinearities~\cite{note}, especially considering the 
high degree of experimental control achieved over the OPO. We also note 
that the above-threshold OPO entanglement renders it a possible device 
for such tasks as a tripartite teleportation network~\cite{AokiNature}. 
Moreover, it allows distribution of quantum information through three 
modes of very different frequencies, a topic that is attracting growing 
attention~\cite{entOPO,gisinnature,prlaustraliens}. This is of practical 
interest, since high efficiency photodetectors are only available in 
limited ranges of the electromagnetic spectrum. Frequency-tunable quantum 
information will also be very useful for light-matter interfaces in 
quantum networks.

We thank J\'anos Bergou for discussions and encouragement. Laboratoire 
Kastler-Brossel, of the Ecole Normale Sup\'{e}rieure and the Universit\'{e} 
Pierre et Marie Curie, is associated with the CNRS (UMR 8552). This work 
was supported by the program CAPES - COFECUB and the Brazilian agencies 
FAPESP and CNPq (\emph{Instituto do Mil\^enio de Informa\c c\~ao Qu\^antica}).



\begin{thebibliography}{99}
\bibitem{note} 
It should be noted, however, that such proposals may lead to even higher orders of 
entanglement, if the pump mode is taken into account.
\end{thebibliography}
\end{document}